\documentclass[3p,preprint]{elsarticle}
\makeatletter
\def\ps@pprintTitle{%
  \let\@oddhead\@empty
  \let\@evenhead\@empty
  \def\@oddfoot{\reset@font\hfil\thepage\hfil}
  \let\@evenfoot\@oddfoot
}
\makeatother

\biboptions{sort&compress}

\usepackage{latexsym}
\usepackage{amsfonts}
\usepackage{amsmath}
\usepackage{amssymb}
\usepackage{mathrsfs}
\usepackage{color}
\usepackage{graphicx}
\usepackage{mathptmx}      
\usepackage{bm}
\usepackage{enumerate}
\usepackage{subfigure}
\usepackage{listings}
\usepackage{grffile}
\usepackage{multirow}
\usepackage{qcircuit}
\usepackage{mathtools}
\usepackage[normalem]{ulem}
\usepackage{pdfpages}
\usepackage{float}
\usepackage{lineno}

\usepackage{xcolor}
\usepackage{hyperref}
\hypersetup{
  colorlinks=true,
}

\definecolor{DarkGreen}{RGB}{0,100,0}

\DeclareRobustCommand\openone{\leavevmode\hbox{\small1\normalsize\kern-.33em1}}

\newcommand\ba{\mathbf{a}}
\newcommand\bb{\mathbf{b}}
\newcommand\bc{\mathbf{c}}

\newcommand\bx{\mathbf{x}}

\newcommand*\widebar[1]{%
   \hbox{%
     \vbox{%
       \hrule height 0.5pt 
       \kern0.5ex
       \hbox{%
         \kern-0.1em
         \ensuremath{#1}%
         \kern-0.1em
       }%
     }%
   }%
}

\newcommand\Cac{1}
\newcommand\Cad{2}
\newcommand\Cbc{3}

\newcommand\TRIP{{\Gamma}}

\newcommand{\LHS}{\mathrm{LHS}}
\newcommand{\RHS}{\mathrm{RHS}}

\newif\ifOLD\OLDfalse
\newif\ifOLD\OLDtrue
\newcommand{\REP}[2]{\ifOLD{{#1}}\else{{\color{black}{#2}}}\fi}
\newif\ifKEEP\KEEPfalse
\newif\ifKEEP\KEEPtrue

\journal{Annals of physics}

\begin{document}
\definecolor{urlorange}{HTML}{ff8000}
\hypersetup{
    colorlinks=true,
    citecolor=red,
    linkcolor=blue,
    filecolor=darkgreen,
    urlcolor=urlorange,
    pdftoolbar=false,
    pdfmenubar=false,
    pdftitle={Foreign exchange rates can violate Bell inequalities},
    }

\begin{frontmatter}

\title{Can foreign exchange rates violate Bell inequalities?}

\cortext[cor1]{Corresponding author: deraedthans@gmail.com\\
Expanded version of the paper published in Ann. Phys.: \url{https://doi.org/10.1016/j.aop.2024.169742}}
\address[FZJ]{J\"ulich Supercomputing Centre, Institute for Advanced Simulation, Forschungzentrum J\"ulich, D-52425 J\"ulich, Germany}
\address[RAD]{Radboud University, Institute for Molecules and Materials, Heyendaalseweg 135, 6525AJ Nijmegen,  Netherlands}
\address[FRA]{Modular Supercomputing and Quantum Computing, Goethe University Frankfurt, Kettenhofweg 139, 60325 Frankfurt am Main, Germany}
\address[RWTH]{RWTH Aachen University, D-52056 Aachen, Germany}

\author[FZJ]{Hans De Raedt\corref{cor1}}
\author[RAD]{Mikhail I. Katsnelson}
\author[FRA]{Manpreet S. Jattana}
\author[FZJ]{Vrinda Mehta}
\author[FZJ]{Madita Willsch}
\author[FZJ]{Dennis Willsch}
\author[FZJ,RWTH]{Kristel Michielsen}
\author[FZJ]{Fengping Jin}

\begin{abstract}
The analysis of empirical data through model-free inequalities leads to the conclusion
that violations of Bell-type inequalities by empirical data
cannot have any significance unless one believes that the universe operates according to the rules of a mathematical model.
\end{abstract}

\date{\today}

\end{frontmatter}
\clearpage
\tableofcontents

\section{Introduction}\label{section0}

The issue of ``violation of Bell inequalities'' is at the center of many heated discussions about
the foundations and interpretations of quantum physics.
The huge interest in the subject is reflected by the award of the
Nobel Prize in physics 2022 ``for experiments with entangled photons, establishing the violation of Bell inequalities and
pioneering quantum information science''.
Remarkably and surprisingly, there is still no consensus about what such violations actually {\bf imply}~\cite{BELL64,PEAR70,PENA72,FINE74,FINE82,FINE82a,FINE82b,MUYN86,KUPC86,BRAN87,JAYN89,BROD89,BROD93,PITO94,FINE96,KHRE09z,
SICA99,BAER99,HESS01a,HESS01b,HESS05,ACCA05,KRAC05,SANT05,
KUPC05,MORG06,KHRE07,ADEN07,Khrennikov2008,NIEU09,MATZ09,KARL09,KHRE09,GRAF09,KHRE11,NIEU11,Brunner2014,HESS15,KUPC16z,KUPC17,HESS17a,NIEU17,
Adenier2017,Khrennikov2018,Drummond2019,Lad2020,Blasiak2021,Cetto2021,Lad2022,Kupczynski2024}.

Recently, we presented new {\bf model-free} inequalities that
reduce to Bell-like inequalities\footnote{
We use this term to refer to Bell's inequality involving three correlations~\cite{BELL64},
the Clauser-Horn-Shimony-Holt (CHSH)~\cite{CLAU69}, Clauser-Horn~\cite{CLAU74}, and all other inequalities that
directly follow from Bell's model for the EPRB thought experiment.}
in a very special case~\cite{RAED23}.
These inequalities put constraints on certain linear combinations of correlations of the (two-valued) data
and, very importantly, are independent of the model that one imagines to have produced the data.
In this paper, we add a new model-free inequality to the family of model-free inequalities presented
in Ref.~\cite{RAED23} and use the new inequality to analyze foreign exchange data.
The model-independent character of the inequalities that we derive, and not the Bell-type inequalities themselves,
provide the appropriate background for discussing
the ``implications'' (whatever they are) of violations of Bell-type inequalities {\bf by empirical data}.
An important point of these model-free inequalities is that their derivation
only exploits elementary arithmetic properties of the data sets and do not refer to physics
at all.
They equally apply to any kind of discrete data~\cite{RAED23}.

Relating these model-free inequalities to physics involves making additional assumptions about
the model that one imagines to have produced the data.
We scrutinize the latter, subtle point by analyzing data of ``non-physical'' origin, namely,
publicly available
\href{https://www.kaggle.com/datasets/brunotly/foreign-exchange-rates-per-dollar-20002019}{
foreign exchange data~\cite{Ferreira24}}, see section~\ref{section1}.

In section~\ref{section2}, we present a new model-free inequality involving only three correlations
and in section~\ref{section3} we use the foreign exchange data to search for violations of the Bell-type inequality
for {\bf empirical data}, obtained as a special case of the model-free inequality.
The main conclusion of this analysis is the following (see section~\ref{section3} for a detailed account).
Computing correlations by considering the {\bf whole} data set, elementary arithmetic dictates that
there can be no violation of the Boole inequality~\cite{BO1862}, a predecessor
of the Bell inequality involving three correlations~\cite{BELL64,BELL93}.
However, dividing (with respect to the time of the transactions) the data set in three equal parts,
the Boole-Bell inequality can be violated (but the corresponding model-free inequality cannot).
Clearly, the observed violation merely reflects {\bf our choice} of selecting data rather than the
``reality'' which, in the present example, is the {\bf complete} data set.
The observation that, depending on the grouping of foreign exchange data,
the analysis can lead to very different conclusions (e.g., violation versus
no violation), is reminiscent of \href{https://en.wikipedia.org/wiki/Simpson's_paradox}{Simpson's paradox~\cite{GRIM01}}.
In this case, the word ``paradox'' does not mean that there is a contradiction involved
but rather emphasizes that viewing the same data in different ways can lead to very different conclusions.

All the Einstein-Podolsky-Rosen-Bohm (EPRB) {\bf laboratory} experiments reporting violations
of Bell-type inequalities that we are aware of employ at least one mechanism for selecting the data from which the correlations are computed.
Thus, potentially they all fall victim to Simpson's paradox.
Most EPRB experiments~\cite{KOCH67,CLAU78,ASPE82b,Kiess1993,WEIH98,CHRI13}
use time-coincidence windows to select pairs of photons.
Other EPRB experiments~\cite{HENS15,GIUS15,SHAL15} use voltage thresholds to classify a detection event
as being the arrival of a photon or as something else.
The sets of empirical data thus obtained can never violate the model-free inequalities~\cite{RAED23}.
Furthermore, unless {\bf all} the cited experiments generate these data such that
they can be reshuffled to form triples and quadruples (which is extremely unlikely),
there is no mathematically sound argument why {\bf these experimental data}
should satisfy any of the Bell-type inequalities~\cite{RAED23}.
As in the case of the foreign exchange data, a violation of Bell-type inequalities
by experimental data merely reflects the properties of the process, {\bf chosen by the experimenter},
to select groups of data, not an intrinsic property of the data themselves.
For instance, the violation of the CHSH inequality by the data of the EPRB experiment by Weihs et al.~\cite{WEIH98}
smoothly changes into a non-violation by increasing the time coincidence window (see Fig.~5 in Ref.~\cite{RAED23}),
a clear case of conclusions depending on viewing the same data differently, recall Simpson's paradox.
For a discussion of the physics aspects of this phenomenon, see section 9 in Ref.~\cite{RAED23}.

The preceding discussion is ``model-free''. Within this framework, one can only prove Bell-type inequalities
if the data satisfy what Boole called ``conditions of possible experience''~\cite{BO1862},
that is if the data derives, without reshuffling, from triples, quadruples~\cite{SICA99}, etc., conditions which are
highly unlikely to be satisfied in any EPRB laboratory experiment (but easily satisfied in computer experiments).
Clearly, in the absence of a proof that Bell-type inequalities exist for general {\bf empirical data},
no conclusion can be drawn from a violation of one of them.

For violations of Bell-type inequalities by experimental data to have any relevance for physics,
it is essential to introduce models that one imagines to be able to produce the data
and derive inequalities from these models.
The simplest but fairly general model for EPRB experiments is undoubtedly the one introduced
by Bell~\cite{BELL64,BELL93}.
Bell's model almost trivially yields Bell-type inequalities which are used to prove Bell's theorem~\cite{BELL93},
stating that Bell's model can never reproduce the full functional form of
the correlation of a quantum system in the singlet state (see sections~\ref{section4} and~\ref{section5}).
As discussed in section~\ref{section4}, Bell's theorem is very important for the foundations of quantum theory.

However, as explained in section~\ref{section6},
these Bell-type inequalities are derived within a particular mathematical model (Bell's model)
and therefore a violation of one or more of them {\bf by empirical data}
only implies that this model cannot serve as a description of the data,
as in the case of our example of the foreign exchange data.

In summary, feeding empirical data from {\bf any} experiment into a Bell-type inequality and observing a violation only
implies that Bell's model does not apply to the case at hand.
In particular, the conclusions must be that Bell's model fails
to describe how the data of actual EPRB experiments are collected and analyzed and that it is necessary
to develop other, better models.
As a matter of fact, a straightforward extension of Bell's model which accounts for the data selection process
can, in the appropriate limit, {\bf exactly} reproduce the results of a quantum system in the singlet state
(see Ref.~\cite{RAED23}, section 11.5 and references therein).
Other conclusions than the two just mentioned constitute logical fallacies, see section~\ref{section7}.

\section{Analysis of foreign exchange rates}\label{section1}

The reader may wonder why the publicly available
\href{https://www.kaggle.com/datasets/brunotly/foreign-exchange-rates-per-dollar-20002019}{
foreign exchange data}
are going to be analyzed by the procedure described in this section.
Comparing this procedure to the one used to analyze data obtained by performing EPRB experiments,
\REP{which is explained in Fig.~\ref{eprbidea}}{see Ref.~\cite{RAED23}(section 3) for a detailed explanation},
it becomes clear that these two procedures are the same.
However, there is no need to be familiar with these experiments to understand this
procedure and the conclusions drawn from it.
All that is necessary to know now is that changes of the exchange rates
have to be digitized (means mapped onto $\pm1$) and that the quantities of interest are the correlations
between the two-valued representation of changes in different exchange rates.

The raw data set contains the exchange rates of the currencies of twenty-two  different countries
relative to the US Dollar, starting on 3 January 2000 and ending on 31 December 2019~\cite{Ferreira24}.
On some days (e.g. Christmas) there is no trading and therefore also no data.
After removing the ``no data'' records, the data set contains
5015 records of twenty-two foreign exchange rates.

In detail, the procedure to calculate correlations is as follows.
\begin{enumerate}
\item
Read the raw ratio data from the file {\sl Foreign\_Exchange\_Rates.csv}
(\href{https://www.kaggle.com/datasets/brunotly/foreign-exchange-rates-per-dollar-20002019}{downloadable here}),
skipping the ``no data'' records and store the floating point data in an array of dimensions (5015,22).
\item
Compute the forward (in time, that is record-wise) differences of these ratios
and store the floating point results in an array of dimensions (5014,22).
\item
Store the sign of all these differences in an integer array of dimensions (5014,22) as values $\pm1$.
These data will be referred to as foreign exchange data in the following.
\item
Divide the $\pm1$ data set into three equal parts of $N=1671$ records (dropping one record),
and denote the parts by ${\cal D}_s$ where the subscript $s=1,2,3$ labels the data set
and the data contained in them.
\item
Compute the correlations
\begin{equation}
C_s(A,B) = \frac{1}{N}\sum_{i=1}^{N} A_{s,i} B_{s,i}
\;,
\label{CORR0}
\end{equation}
where $A$ and $B$ each symbolize one of the twenty-two currencies and $A_{s,i}=\pm1$, $B_{s,i}=\pm1$.
Excluding the trivial correlation
$C_s(A,A)=1$ and noting that $C_s(A,B)=C_s(B,A)$ for $s=1,2,3$,
this procedure yields $3\times 21 \times 22/2 = 693$ different correlations.
\end{enumerate}

\section{Model-free inequalities}\label{section2}
The prime focus of this paper is on the conclusions that can be drawn
from violations of inequalities on certain
linear combinations of the correlations $C_s(A,B)$, $C_{s'}(A',B')$, etc.
As explained next, without knowing the actual values of these correlations
or without assuming a particular model for the process that generates the data,
elementary arithmetic alone already yields nontrivial inequalities that can never
be violated by data.

As the $C_s(A,B)$'s are averages of $\pm1$ values, it follows immediately that $-1\le C_s(A,B)\le 1$
for $s=1,2,3$ and all pairs $(A,B)$ of currencies.
Then, obviously, $-2\le C_s(A,B)+ C_{s'}(A',B')\le 2$
where $(A,B)$ and $(A',B')$ denote any two pairs of currencies and $s,s'=1,2,3$.
Then what about $C_1(A,B)+ C_2(A',B')+ C_3(A'',B'')$, for instance?

Let $(x,y,x',y',x'',y'')$ stand for
any of the sextuples $(A^{\phantom{'}}_{1,i},B^{\phantom{'}}_{1,i},A'_{2,i},B'_{2,i},A''_{3,i},B''_{3,i})$.
If $x,y,x',y',x'',y''=\pm1$ it follows immediately that $-3\le xy + x'y' + x''y'' \le 3$
and therefore, $-3\le C_1(A,B)+ C_2(A',B')+ C_3(A'',B'')\le 3$. But can one find stricter bounds?

Recall that a basic property of a sum of numbers is that the order in which
the numbers are added is irrelevant.
This elementary arithmetic fact and this fact alone
can reduce the values of contributions to $C_1(A,B)+ C_2(A',B')+C_3(A'',B'')$.
To see this, use the freedom to reshuffle the contributions to
$C_2(A',B')$ and $C_3(A'',B'')$ and consider the expression
$A^{\phantom{'}}_{1,i}\,B^{\phantom{'}}_{1,i}+ A'_{2,j}\,B'_{2,j}+ A''_{3,k}\,B''_{3,k}$.
Further assume that for a given $i$, it is possible to find at least one pair $(j,k)$ such that
$A'_{2,j}=A^{\phantom{'}}_{1,i}$, $B'_{2,j}=B''_{3,k}$ and $A''_{3,k}=B{\phantom{'}}_{1,i}$.
In this particular case, the sextuple $(A^{\phantom{'}}_{1,i},B^{\phantom{'}}_{1,i},A'_{2,j},B'_{2,k},A''_{3,k},B''_{3,k})$
derives from the triple $(x=A^{\phantom{'}}_{1,i},y=B^{\phantom{'}}_{1,i},z=B'_{2,j})$
and its contribution to $C_1(A,B)+ C_2(A',B')+ C_3(A'',B'')$
is given by $xy + xz + yz$.
It is easy to verify that $-1\le xy + xz + yz \le 3$ by
considering all eight combinations of $x,y,z=\pm1$.
Thus, if by reshuffling the data it becomes possible to reduce sextuples to triples,
the contributions of sextuples is bounded from below by minus one instead of minus three.
In other words, reshuffling can reduce the contributions to linear combinations of the three correlations.

\REP{In~\ref{TRIPLES} we exploit this elementary property of integer arithmetic to }{By
a straightforward extension of the proof given in Ref.~\cite{RAED23}(appendix B), we can prove }that in general
\begin{align}
\vert C_{1}(A,B) \pm C_{2}(A',B')\vert&\le 3-2\TRIP \pm C_{3}(A'',B'')
\;,
\label{TRIP0}
\end{align}
or, equivalently
\begin{align}
\vert C_{1}(A,B) \pm C_{2}(A',B')\vert\mp C_{3}(A'',B'')-1&\le 2(1-\TRIP)
\;,
\label{TRIP0a}
\end{align}
where $\TRIP$ denotes the maximum fraction of triples that one can identify
by reshuffling the data in ${\cal D}_{s}$ with $s=2,3$.
\REP{In \ref{MAXTRIP} it is shown how to compute $\TRIP$ from the data}{
$\TRIP$ can be computed by a slightly modified version of the procedure
described in Ref.~\cite{RAED23}(appendix B)}.
Note that Eq.~(\ref{TRIP0}) is equivalent to
$\vert C_{1}(A,B) \pm C_{3}(A'',B'')\vert\le 3-2\TRIP \pm C_{2}(A',B')$ or
$\vert C_{3}(A'',B'') \pm C_{2}(A',B')\vert\le 3-2\TRIP \pm C_{1}(A,B)$,
\REP{as shown in~\ref{BASIC}}{which follow directly from Eq.~(N.6) in Ref.~\cite{RAED23}}.

If and only if all the $N$ pairs of records in the three data sets
${\cal D}_{\Cac}$, ${\cal D}_{\Cad}$, and ${\cal D}_{\Cbc}$
can be reshuffled to create $N$ triples, the fraction of triples $\TRIP=1$.
In this particular case, Eq.~(\ref{TRIP0})
takes the form of the Boole inequality~\cite{BO1862}
\begin{align}
\vert C_{1}(A,B) \pm C_{2}(A,B')\vert\le 1 \pm C_{3}(B,B')
\;,
\label{BELL0}
\end{align}
written here in a different but equivalent form than Boole did.

The inequality Eq.~(\ref{TRIP0}) (for $0\le\TRIP\le1$)
holds for any set of sextuples
${\cal D}=\{(A_{1,i},B_{1,i},A'_{2,i},B'_{2,i},A''_{3,i},B''_{3,i})\,|\,i=1,\ldots,N\}$,
irrespective of how the data was obtained or generated.
It is therefore {\bf model free}, meaning that inequality Eq.~(\ref{TRIP0}) holds
for data, regardless of any imaginary model that is believed to have produced these data.
Model-free inequalities, involving four correlations and
reducing to the Clauser-Horn-Shimony-Holt~\cite{CLAU74,BELL93} and Clauser-Horn~\cite{CLAU69}
inequalities in the exceptional case that all octuples of data can be reshuffled to form
quadruples, are given in Ref.~\cite{RAED23}.

\section{Application of the model-free inequalities to the foreign exchange data}\label{section3}

It is clear that Eq.~(\ref{TRIP0}), being the result of basic arithmetic only, can never be violated if $\Gamma$ is chosen as defined.
With regard to Bell-type inequalities, to be discussed in section~\ref{section6}, the main question of interest is
``can certain combinations of the foreign exchange data violate Eq.~(\ref{TRIP0}) with $\TRIP=1$?''
Note that if these data were uniformly random, $\TRIP=1$ up to
statistical fluctuations\REP{, see ~\ref{MAXTRIP} for an example}{.
The proof of this fact is similar to the one given in Ref.~\cite{RAED23}(appendix B).}

Computing the pairwise correlations of all foreign exchange rates according to the procedure outlined in section~\ref{section1},
there are 85 out of 18480 possible combinations of triples of foreign currencies
that violate at least one of Bell-like inequalities
\begin{align}
\vert C_{1}(A,B) \pm C_{2}(A,C)\vert\mp C_{3}(B,C)-1 &\le0
\;.
\label{TRIP1}
\end{align}
The maximum value of $\vert C_{1}(A,B) + C_{2}(A,C)\vert - C_{3}(B,C)-1 = 0.17$
with $A$ representing the Euro, $B$ the Swiss Franc, and $C$ the Danish Krone.
The average values of the $A$'s, $B$'s, and $C$'s are at most $0.012$, and
$C_{1}(A,B)=0.80$, $C_{2}(A,C)=0.96$, and $C_{3}(B,C)=0.59$.
Calculating the maximum number of triples by solving the minimization problem
\REP{(see~\ref{MAXTRIP} for a description of the procedure)}{by a
slightly modified version of the procedure described in Ref.~\cite{RAED23}(appendix B)}
yields $2(1-\TRIP)=0.17$,
not only in agreement with Eq.~(\ref{TRIP0}) (with the plus sign) but also signaling that the left hand side
of Eq.~(\ref{TRIP0a}) is equal to the bound $2(1-\TRIP)$.

The maximum value of $\vert C_{1}(A,B) - C_{2}(A,C)\vert + C_{3}(B,C)-1 = 0.23$
where in this case, $A$ represents the Mexican Peso, $B$ the Euro, and $C$ the Danish Krone.
The average values of the $A$'s, $B$'s, and $C$'s are at most $0.021$,
$C_{1}(A,B)=-0.039$, $C_{2}(A,C)=0.25$, and $C_{3}(B,C)=0.94$.
Also in this case, the value of the left hand side of Eq.~(\ref{TRIP0a}) (with the minus sign)
is the same as $2(1-\TRIP)=0.23$.

If, for simplicity, it is assumed that the standard deviation on the correlations is
approximately given by $1/2\sqrt{N}=1/2\sqrt{1671}\approx0.012$, the foreign exchange data
shows violations of more than 10 standard deviations in the two cases mentioned earlier.

Finally, it should be mentioned that using
$C(A,B) = (3N)^{-1}\sum_{s=1}^{3}\sum_{i=1}^{N} A_{s,i} B_{s,i}$ to compute the correlations
(that is without breaking up each of the whole data sets in three parts),
and repeating the analysis never produces
violations of $\vert C(A,B) \pm C(A,C)\vert\mp C(B,C)-1 \le0$.
This is to be expected because in this case, the data for the $A$'s, $B$'s, and $C$'s
form triples ($\TRIP=1$) and then, as already shown by Boole~\cite{BO1862}, there can be no violation.

\section{Mathematical models: importance of Bell's theorem}\label{section4}

In brief, Bell proposed to model the correlations of an EPRB thought experiment
\REP{(see~\ref{EPRB}) }{}by~\cite{BELL71,BELL93}
\begin{eqnarray}
{\cal C}(\ba,\bb)=\int {A}(\ba,\lambda){B}(\bb,\lambda)\,\mu(\lambda)\,d\lambda
\;,\;
|{A}(\ba,\lambda)|\le1
\;,\;
|{B}(\bb,\lambda)|\le1
\;,\;0\le\mu(\lambda)
\;,\;\int \mu(\lambda)\,d\lambda=1
\;,
\label{IN0}
\end{eqnarray}
where ${A}(\ba,\lambda)$ and ${B}(\bb,\lambda)$ are
mathematical functions of the conditions $\ba$ and $\bb$, respectively,
and the common variable $\lambda$ denoting an arbitrary set of ``hidden'' variables.
Bell gave a proof that ${\cal C}(\ba,\bb)$
cannot arbitrarily closely approximate the correlation $-\ba\cdot\bb$ {\it for all}
unit vectors $\ba$ and $\bb$~\cite{BELL64}.
According to Bell himself (see Ref.~\cite{BELL93}(p.65)), {\bf this is the theorem.}

{\bf Within the universe of mathematical models}, Bell's theorem is of great importance.
Apparently not well-known seems to be the fact that the theorem excludes a probabilistic
description of the Stern-Gerlach experiment with spin-1/2 particles
in terms of the model Eq.~(\ref{IN0})~\cite{RAED23}.
In this particular case, the quantum-theoretical description goes in terms of a single particle.
Therefore all implications pertaining to physics, other than the one just mentioned,
drawn from a violation of a Bell-type inequalities become void.
Very well-known is the fact that the theorem excludes
all models of the type Eq.~(\ref{IN0}) as possible
candidates for describing a quantum system of two spin-1/2 objects
for which in a certain case ${\cal C}(\ba,\bb)=-\ba\cdot\bb$.
By far the most important general consequence of Bell's theorem is that there is no hope for
recovering {\bf all} the results of quantum theory by expressions of the kind Eq.~(\ref{IN0}),
that is by averaging (the integration over $\lambda$ with
probability density $\mu(\lambda)$) over an ensemble of ``classical'' physics models
formulated in terms of scalar functions with values in the interval $[-1,+1]$.

Equation~(\ref{IN0}) generalizes the idea of separation of variables, that is
the idea that a scalar function of two variables ($\ba$ and $\bb$) can, in particular cases, be
written as (the integral over) a product of scalar functions, each of which depend on one of these variables only.
{\bf Within the universe of mathematical models},
there is absolutely no valid argument for restricting the search for models
yielding correlations ${\cal C}(\ba,\bb)=-\ba\cdot\bb$
to the domain of models formulated in terms of scalar functions (as in Eq.~(\ref{IN0})).
In fact, as is well-known, quantum theory provides such a separated model in terms of Pauli spin matrices~\cite{RAED23}.
Moreover, with a suitable definition of the notion of locality, the quantum-theoretical model
can also be argued to exhibit locality~\cite{Howard1985,Deutsch2000,RaymondRobichaud2021,Bedard2021}.
Another possibility is to resort to non-Diophantine arithmetics~\cite{Czachor2021}.

\section{Mathematical models: implications of violating Bell-type inequalities}\label{section5}

From Eq.~(\ref{IN0}),
\REP{Eq.~(\ref{BASIC3}),}{$|ac \pm ac| \le1\pm bc$ for $-1\le a,b,c \le1$,}
and the application of the triangle inequality,
it follows directly that
\begin{align}
\vert {\cal C}(\ba,\bb) \pm {\cal C}(\ba,\bc)\vert\le 1\pm {\cal C}'(\bb,\bc)
\;,
\label{IN1}
\end{align}
where
\begin{eqnarray}
{\cal C}'(\bb,\bc)=\int {B}(\bb,\lambda){B}(\bc,\lambda)\,\mu(\lambda)\,d\lambda
\;.
\label{IN0a}
\end{eqnarray}
Inequality~(\ref{IN1}) will be referred to as a Boole-Bell inequality.
Key arguments in Bell's original proof of his theorem are the assumption of perfect anticorrelation
(meaning ${A}(\bx,\lambda)=-{B}(\bx,\lambda)$ for all $\bx$) and that ${\cal C}(\ba,\bb)=\pm\ba\cdot\bb$.
Bell then shows that Eq.~(\ref{IN1}) can be violated~\cite{BELL64,BELL93}.
For instance, with the choice $\ba=(1,0,0)$, $\bb=(1,1,0)/\sqrt{2}$, and $\bc=\pm(1,-1,0)/\sqrt{2}$,
Eq.~(\ref{IN1}) becomes $\sqrt{2}\le1$, a clear violation.

Always within the realm of mathematical models, the only logically correct
conclusion that one can draw from a violation of Eq.~(\ref{IN1}) is that
the model Eq.~(\ref{IN0}) cannot describe the correlation $\pm\ba\cdot\bb$ for all unit vectors
$\ba$ and $\bb$, which is Bell's theorem.
However, that does not imply that only the model Eq.~(\ref{IN0}) is excluded.

By virtue of Fine's theorem~\cite{FINE82a,FINE82b} (for an alternative proof, see \ref{FINE}),
for a fixed triple of conditions $(\ba,\bb,\bc)$,
a violation of Eq.~(\ref{IN1}) also
excludes the existence of any description in terms of
a normalized nonnegative distribution $f(x,y,z)$
($0\le f(x,y,z) \le 1$, $\sum_{x,y,z=\pm1} f(x,y,z)=1$)
with the property that
\begin{align}
{\cal C}(\ba,\bb) =\sum_{x,y,z=\pm1} xy\,f(x,y,z)\;,\;
{\cal C}(\ba,\bc) =\sum_{x,y,z=\pm1} xz\,f(x,y,z)\;,\;
{\cal C}(\bb,\bc) =\sum_{x,y,z=\pm1} yz\,f(x,y,z)
\;.
\label{IN2}
\end{align}
As a matter of fact, Fine's theorem shows much more, namely that
the existence of a normalized nonnegative distribution $f(x,y,z)$
is equivalent to the Boole-Bell inequalities being satisfied
(for a technically precise statement, see \ref{FINE}).

From a violation of Eq.~(\ref{IN1}) one can, if one wishes to do so, draw
the conclusion that the correlations cannot be obtained from the model Eq.~(\ref{IN0})
but Fine's theorem leaves no room for
interpretations of violations of a Boole-Bell inequality other than the
mathematical equivalence.

It is important to mention here that Fine's theorem has no implications for Bell's theorem.
Furthermore, the existence of a normalized nonnegative distribution $f(x,y,z)$
{\bf for a fixed triple of conditions $(\ba,\bb,\bc)$}
does not imply that there exists a probabilistic description
of the EPRB experiment in terms of a trivariate, simply
because one cannot perform one EPRB experiment conditioned on the three settings $(\ba,\bb,\bc)$.
In order to do so, one would needs to perform an extended EPRB experiment~\cite{RAED20a,RAED23}.
In this case, it is impossible to violate any of the Bell-type inequalities~\cite{RAED20a,RAED23}.

\section{Empirical data: irrelevance of violating Bell-type inequalities}\label{section6}

As shown in section~\ref{section1}%
\ifKEEP
and~\ref{TRIPLES}
\fi,
correlations computed from two-valued data must {\bf always} satisfy the inequality Eq.~(\ref{TRIP0})
with $\TRIP$ being the value of the fraction of triples.
The Boole-Bell inequality {\bf for empirical data} Eq.~(\ref{BELL0}) is recovered in the exceptional
case that $\TRIP=1$, the {\bf only} case in which one can prove the Boole-Bell inequality for empirical data.
The same is true for the Clauser-Horn-Shimony-Holt~\cite{CLAU74,BELL93} and Clauser-Horn~\cite{CLAU69}
inequalities~\cite{RAED23}.
For empirical data, the mathematical proof of these two inequalities can only be carried out if the empirical data can be reshuffled
to form quadruples~\cite{RAED23}.

The violation of Bell-type inequalities by data from EPRB experiments and the conclusions
drawn from it are regarded as a landmark in the development of modern quantum technologies.
But what then with the violation of Boole-Bell inequalities by the foreign exchange rate data?
The idea that this would be a consequence of ``quantum or non-classical physics at work'' sounds strange, to put it mildly.
Clearly, the logic, arguments and concepts (such as locality as encapsulated by Eq.~(\ref{IN0})) that are being used
in contemporary quantum physics to interpret violations of Bell-type inequalities need to be scrutinized further.

\section{Conclusion}\label{section7}

The discussion that follows uses the three-correlations case as an example.
The same arguments and conclusions almost trivially extend to the four-correlations case as well~\cite{RAED23}.
The results presented in the earlier sections can be summarized as follows:
\begin{enumerate}
\item
Correlations of empirical data cannot violate the model-free inequality Eq.~(\ref{TRIP0}).
\item
Correlations of empirical data can violate inequality Eq.~(\ref{BELL0})
when not all sextuples of data can be rearranged to form triples (i.e., $\TRIP<1$).
Such a violation has no meaning (other than that not all the triples can form sextuples)
because empirical data should comply with Eq.~(\ref{TRIP0}), not with Eq.~(\ref{BELL0}),
as exemplified by the analysis of the foreign exchange data.
\item
The proof of inequality Eq.~(\ref{IN1}) assumes that the correlation is given by model Eq.~(\ref{IN0}).
If a properly discretized version of model Eq.~(\ref{IN0}) is used as a (computer) model
for producing synthetic data, this model generates data that can be reshuffled into a set of triples~\cite{RAED23},
that is $\TRIP=1$~\cite{RAED23}.
\item
A violation of the Bell-type inequality Eq.~(\ref{BELL0})
by empirical data rules out the model Eq.~(\ref{IN0}) as a potential
candidate for {\bf describing} the data.
If the empirical data cannot be described by Bell's model Eq.~(\ref{IN0}), the model has to be rejected or extended.
\item
Unless one subscribes to the idea that empirical data are generated by
``mathematical rules governing the universe'',
the role of any mathematical model is limited to {\bf describing} empirical data.
If the model description fails (e.g., because a model-dependent inequality is violated),
the proper action, practiced in most fields of science,
would be to try improving the model, not to philosophize about the premises that went into its formulation.
\end{enumerate}

From (1--5), it follows that it is only by adopting the view that a mathematical model such as Eq.~(\ref{IN0})
{\bf is} a reality existing in the world in which we live that it may become possible
to interpret the premises that led to the formulation of Eq.~(\ref{IN0}) as genuine properties of ``nature''.
As it is unknown whether ``mathematics rules the universe'',
it is perhaps more apt to analyze data without relying on the premise that it does.

\section*{Acknowledgments}
The work of M.I.K. was supported by the European Research Council (ERC) under the European Union's
Horizon 2020  research and innovation programme, grant agreement 854843 FASTCORR.
V.M., D.W., and M.W. acknowledge support from the project J\"ulich UNified Infrastructure for Quantum computing (JUNIQ)
that has received funding from the German Federal Ministry of Education and Research (BMBF)
and the Ministry of Culture and Science of the State of North Rhine-Westphalia.

\appendix
\ifKEEP
\section{Einstein-Podolsky-Rosen-Bohm thought experiment}\label{EPRB}
Figure~\ref{eprbidea} shows the layout of the EPRB thought experiments and its caption
describes the procedure used to collect the data.
\begin{figure}[H]
\centering
\includegraphics[width=0.90\hsize]{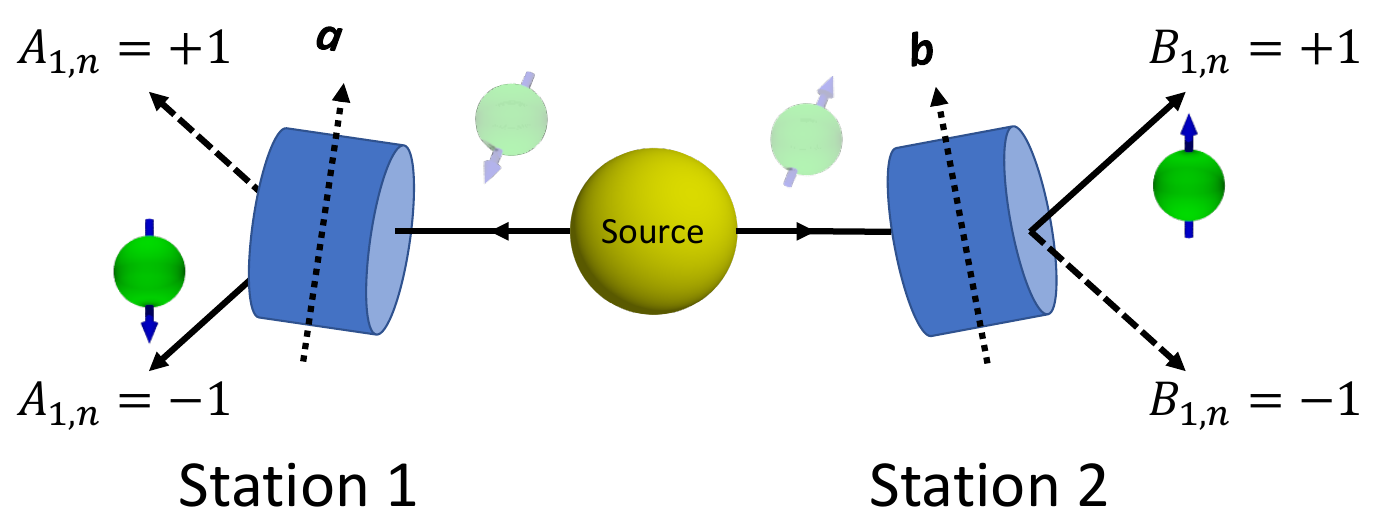}
\caption{
Conceptual representation of the Einstein-Podolsky-Rosen thought experiment~\cite{EPR35}
in the modified form proposed by Bohm~\cite{BOHM51}.
A source produces pairs of particles.
The particles of each pair carry opposite magnetic moments implying that
there is a correlation between the two magnetic moments of each pair leaving the source.
The magnetic field gradients of the Stern-Gerlach magnets (cylinders)
with their uniform magnetic field component
along the directions of the unit vectors $\ba$ and $\bb$ divert each incoming
particle into one of the two, spatially separated directions labeled by $+1$ and $-1$.
The pair $(\ba,\bb)$ determines the conditions, to be denoted by the subscript ``$1$'',
under which the data $(A_{1,n},B_{1,n})$ is collected.
The values of $A_{1,n}$ and $B_{1,n}$ correspond to the labels of the
directions in which the particles have been diverted.
The result of this experiment is the set of data pairs
${\cal D}_{1}=\{ (A_{1,1},B_{1,1}),\ldots, (A_{1,N},B_{1,N})\}$
where $N$ denotes the total number of pairs emitted by the source.
}
\label{eprbidea}
\end{figure}

\section{Proof of the model-free inequality for triples of correlations of discrete data}\label{TRIPLES}

In mathematics, an $n$-tuple is an ordered list of $n$ elements, denoted by $(A,B,\ldots)$.
$(A,B)$ is called a pair, implicitly assuming it is ordered, a 2-tuple.
Similarly, the lists $(A,B,C)$ and $(A,B,C,D)$ are
referred to as a triple (3-tuple) or quadruple (4-tuple),
and an ordered list of 6 (8) elements is a sextuple (octuple).
Discrete data are finite collections of data, each data item only taking a limited number of values.
Each of these items comes either directly in the form of (ratios of) integers or can be mapped
one-to-one onto the latter.

This appendix gives the details of the proof of a rigorous bound on certain combinations of correlations
between pairs of data items making up sextuples.
Essential thereby is that there is no assumption on how the data came into existence,
that is, the resulting inequality is model free.
Any procedure that yields sextuples of {\bf discrete} data,
which without loss of generality can always
be thought of as being rescaled to lie in the interval $[-1,+1]$,
must satisfy this inequality.

If the data items take values $\pm1$ only, the model-free inequality
contains, {\bf as a special case}, a model-free inequality first derived by Boole~\cite{BO1862}.
This exceptional case was rediscovered by Bell~\cite{BELL64} in the context of a mathematical model
for the EPRB thought experiment with the supplementary assumption of perfect anticorrelation.

The $n$th discrete data item obtained under the condition labeled by $s$ is denoted
by e.g., $A_{s,n}$ for $n=1,\ldots,N$.
The subscript $s$ labels the instance only.
In this section, $|A_{s,n}|\le1$, including the case of interest $A_{s,n}=\pm1$.
The presence of $s$ does not, in any way, imply
that there is an underlying mechanism or (mathematical) model that governs
the dependence of $A_{s,n}$ on $s$.

For the derivation of the model-free inequality for triples of correlations of discrete data,
it is convenient to write (without loss of generality) the set of sextuples as
\begin{eqnarray}
{\cal S}&=&\{(A_{\Cac,n},B_{\Cac,n},A_{\Cad,n},B_{\Cad,n},A_{\Cbc,n},B_{\Cbc,n})
\,|\,|A_{s,n}|\le1,|B_{s,n}|\le1;\,s=1,2,3\,;\,n=1,\ldots,N\}
\;.
\label{SEXT0}
\end{eqnarray}
To prepare for the calculation of correlations between pairs of data, we
extract from the set Eq.~(\ref{SEXT0}), the three data sets
\begin{eqnarray}
{\cal D}_{s}&=&\{(A_{s,n},B_{s,n})\,|\,
|A_{s,n}|\le1,|B_{s,n}|\le1\,;\,n=1,\ldots,N\} \quad,\quad s=1,2,3
\;,
\label{DISD0}
\end{eqnarray}
where $N$ is the number of pairs.
Correlations between the discrete data Eq.~(\ref{DISD0}), are computed according to
\begin{eqnarray}
C_{s}=\frac{1}{N}\sum_{n=1}^N A_{s,n}B_{s,n}\quad,\quad s=1,2,3
\;.
\label{DISD1}
\end{eqnarray}

In general, each contribution to the correlations Eq.~(\ref{DISD1})
may take any value in the interval $[-1,+1]$, independent of the values taken by other contributions,
yielding the trivial bounds
\begin{eqnarray}
|C_{i}\pm C_{j}|\le2
\quad,\quad(i,j)\in\{(1,2),(1,3),(2,3)\}
\;.
\label{DISD2}
\end{eqnarray}
Without introducing a specific model for the process that generates the data,
sharp bounds can be derived by identifying the contributions to $C_{1}\pm C_{2}$
which, after suitable reshuffling of the terms, can be brought in the form $xy\pm xz$ (where the discrete
data $x,y,z$ satisfy $\vert x\vert$, $\vert y\vert$, $\vert z\vert$ $\le 1$).
In this case, $\vert xy\pm xz\vert\le 1\pm yz$ holds (see Eq.~(\ref{BASIC3})).
Therefore, for contributions to $C_{1}\pm C_{2}$ which exhibit this ``triple structure'',
the value of $xy\pm xz$ lies in the interval $[-(1\pm yz),1\pm yz]\subseteq [-2,2]$.

First, introduce permutations $P(.)$, ${\widehat P}(.)$, and ${\widetilde P}(.)$
acting on elements of the set $\{1,2,\ldots,N\}$
and rewrite Eq.~(\ref{DISD1}) as
\begin{eqnarray}
C_{\Cac}=\frac{1}{N}\sum_{n=1}^N A_{\Cac,{P}(n)}B_{\Cac,{P}(n)}\;,\;
C_{\Cad}=\frac{1}{N}\sum_{n=1}^N A_{\Cad,{\widehat P}(n)}B_{\Cad,{\widehat P}(n)}\;,\;
C_{\Cbc}=\frac{1}{N}\sum_{n=1}^N A_{\Cbc,{\widetilde P}(n)}B_{\Cbc,{\widetilde P}(n)}\;
\;.
\label{QUAD0}
\end{eqnarray}
As addition is commutative, reordering the terms of the sums does not change the value
of the sums themselves.

For a particular choice of the three permutations
$P$, ${\widehat P}$, and ${\widetilde P}$
acting on elements of the set
$\{1,2,\ldots,N\}$,
let $K$ be the number of triples $(x_k,y_k,z_k)$  such that
\begin{align}
x_k&=A_{\Cac,P(k)}=A_{\Cad,{\widehat P}(k)}
\;,&\;
y_k&=B_{\Cac,P(k)}=A_{\Cbc,{\widetilde P}(k)}
\;,&\;
z_k&=B_{\Cad,{\widehat P}(k)}=B_{\Cbc,{\widetilde P}(k)}
\;,
\label{QUAD3}
\end{align}
for $k=1,\ldots,K$.
Obviously, the value of $K$ depends on the choice of the permutations (the data itself is considered to be fixed).
Let $K_{\mathrm{max}}$ denote the maximum value of $K$ that is
obtained by exploring all possible triples of permutations
$P$, ${\widehat P}$, and ${\widetilde P}$ of $\{1,2,\ldots,N\}$.
The maximum fraction of triples is defined by $\TRIP=K_{\mathrm{max}}/N$.
Obviously, $0\le\TRIP\le1$.

Our choice Eq.~(\ref{QUAD3}) to define triples expresses the notion of ``possible experience''
used by Boole to prove what is now known as a Bell inequality~\cite{BO1862}, or, equivalently
expresses the assumption of counterfactual definiteness implicitly used by Bell in the derivation
of his (three-term) inequality~\cite{BELL64,BELL93}.
Note that choices other than Eq.~(\ref{QUAD3}) may yield different values of the maximum fraction $\Gamma$ of triples.

For each $k=1,\ldots,K_{\mathrm{max}}$, application of the inequality $\vert xy\pm xz\vert\le 1\pm yz$
(see Eq.~(\ref{BASIC3})) yields
\begin{align}
\vert A_{\Cac,{P}(k)}B_{\Cac,{P}(k)}\pm A_{\Cad,{\widehat P}(k)}B_{\Cad,{\widehat P}(k)}\vert
\le 1 \pm B_{\Cac,{P}(k)} B_{\Cad,{\widehat P}(k)} = 1 \pm A_{\Cbc,{\widetilde P}(n)}B_{\Cbc,{\widetilde P}(n)}
\;.
\label{TRI0}
\end{align}
or, written slightly differently
\begin{subequations}
\label{TRI1}
\begin{align}
A_{\Cac,{P}(k)}B_{\Cac,{P}(k)} \pm A_{\Cad,{\widehat P}(k)}B_{\Cad,{\widehat P}(k)}
\mp A_{\Cbc,{\widetilde P}(n)}B_{\Cbc,{\widetilde P}(n)}
&\le 1\;,
\label{TRI1a}
\\
A_{\Cac,{P}(k)}B_{\Cac,{P}(k)} \pm A_{\Cad,{\widehat P}(k)}B_{\Cad,{\widehat P}(k)}
\pm A_{\Cbc,{\widetilde P}(n)}B_{\Cbc,{\widetilde P}(n)}
&\ge -1
\;.
\label{TRI1b}
\end{align}
\end{subequations}
For $k>K_{\mathrm{max}}$, the right hand sides in Eq.~(\ref{TRI1}) have to be replaced by $3$ and $-3$, respectively.

Splitting each of the sums in Eq.~(\ref{QUAD0}) into a sum over $k=1,\ldots,K_{\mathrm{max}}$
and using Eq.~(\ref{TRI1a}) yields
\begin{align}
C_{\Cac} \pm C_{\Cad} \mp C_{\Cbc}&=
\frac{1}{N}\sum_{k=1}^{K_{\mathrm{max}}}
\left(
A_{\Cac,{P}(k)}B_{\Cac,{P}(k)} \pm A_{\Cad,{\widehat P}(k)}B_{\Cad,{\widehat P}(k)}
\mp A_{\Cbc,{\widetilde P}(n)}B_{\Cbc,{\widetilde P}(n)}
\right)
\nonumber \\
&+\frac{1}{N}\sum_{n=K_{\mathrm{max}}+1}^N
\left(
A_{\Cac,{P}(k)}B_{\Cac,{P}(k)} \pm A_{\Cad,{\widehat P}(k)}B_{\Cad,{\widehat P}(k)}
\mp A_{\Cbc,{\widetilde P}(n)}B_{\Cbc,{\widetilde P}(n)}
\right)
\\
&\le \TRIP + 3(1-\TRIP)= 3-2\TRIP
\;.
\label{DISD7}
\end{align}
Similarly, using Eq.~(\ref{TRI1b}) gives
\begin{align}
C_{\Cac} \pm C_{\Cad} \pm C_{\Cbc}\ge -3+2\TRIP
\;.
\label{DISD8}
\end{align}
Combining Eqs.~(\ref{DISD7}) and Eq.~(\ref{DISD8}), yields
\begin{align}
\vert C_{\Cac} \pm C_{\Cad}\vert\le 3-2\TRIP \pm C_{\Cbc}
\;.
\label{DISD9}
\end{align}

Finally, note that if the real numbers $a$, $b$, $c$ and $d$
satisfy the inequalities $|a\pm b|\le d \pm c$,
these numbers also satisfy the inequalities $|a\pm c|\le d \pm b$
and $|c\pm b|\le d \pm a$.
Indeed, more explicitly the inequalities $|a\pm b|\le d \pm c$
read $-d\mp c \le a\pm b\le d \pm c$ from which
$a\mp c\le d\mp b$ and $-d\mp b\le a\pm c$ or $|a\pm c|\le d\pm b$.
Similarly, it follows that $|c\pm b|\le d \pm a$.
Therefore, inequalities Eq.~(\ref{DISD9}) imply that in general
\begin{align}
\vert C_{i} \pm C_{j}\vert\le 3-2\TRIP \pm C_{k}\quad,\quad (i,j,k)\in\bm\pi_3
\;,
\label{TRI10}
\end{align}
where $\bm\pi_3$ denotes the set of all permutations of $(1,2,3)$.

Remark that if the correlations satisfy $\vert C_{i} \pm C_{j}\vert\le 1 \pm C_{k}$
for all $(i,j,k)\in\bm\pi_3$, that is if $\TRIP=1$, it does not follow that the data in  the sets
${\cal D}_{\Cac}$, ${\cal D}_{\Cad}$, and ${\cal D}_{\Cbc}$
can be reshuffled to form triples.
In fact, it is easy to construct simple counter examples.
For instance, if
${\cal D}_{\Cac}=\{(+1,-1),(+1,+1)\}$,
${\cal D}_{\Cad}=\{(-1,-1),(-1,+1)\}$, and
${\cal D}_{\Cbc}=\{(+1,+1),(-1,-1)\}$,
then $C_{\Cac}=C_{\Cad}=0$ and $C_{\Cbc}=1$ and the inequalities
Eq.~(\ref{TRI10}) with $\TRIP=1$
are satisfied but the data in the three sets cannot be reshuffled to form triples.

In summary, the inequalities Eq.~(\ref{TRI10})
hold in general, independent of (any model for) the process that generates the data.
The inequalities Eq.~(\ref{TRI10}) cannot be violated by discrete data (of a real or thought or computer experiment),
unless the mathematical apparatus that is being used is inconsistent (a possibility that is not considered).

\subsection{Computing the maximum number of triples}\label{MAXTRIP}
The maximum number of triples can be calculated by slightly modifying the procedure described
in Ref.~\cite{RAED23} used to compute the maximum number of quadruples.
The key step is to list all possible eight combinations of $A$'s and $B$'s that
form triples and to attach a variable to each of these combinations, as shown in Table~\ref{tab4}.
The numbers of $(A,B)$ pairs in the data sets ${\cal D}_s$ for $s=1,2,3$
are denoted by $n_s(A_s,B_s)$ and, as shown by Table~\ref{tab5},
can be expressed in terms of the $m_i$ in Table~(\ref{tab4}).

\begin{table}[!htp]
\caption{
Lists of all possible combinations of the pairs of data which form triples, written
in a slightly different order to emphasize the triple structure.
Given the data sets ${\cal D}_{\Cac}$, ${\cal D}_{\Cad}$, and ${\cal D}_{\Cbc}$,
the optimization task is to find the numbers $m_i\ge0$ and $u_i\ge0$ for $i=0,\ldots,7$ that
maximize the number of triples subject to 35 constraints (see text).
}
\centering
\begin{tabular}{@{\extracolsep{1cm} } ccccc}
\noalign{\medskip}
\hline\hline\noalign{\smallskip}
& $(B_1,A_1)$  &$(A_2,B_2)$& $(B_3,A_3)$ \\
\hline\noalign{\smallskip}
$m_0   $&  $(+1,+1)$ & $(+1,+1)$ &$(+1,+1)$\\
$m_1   $&  $(-1,+1)$ & $(+1,+1)$ &$(+1,-1)$\\
$m_2   $&  $(+1,+1)$ & $(+1,-1)$ &$(-1,+1)$\\
$m_3   $&  $(-1,+1)$ & $(+1,-1)$ &$(-1,-1)$\\
$m_4   $&  $(+1,-1)$ & $(-1,+1)$ &$(+1,+1)$\\
$m_5   $&  $(-1,-1)$ & $(-1,+1)$ &$(+1,-1)$\\
$m_6   $&  $(+1,-1)$ & $(-1,-1)$ &$(-1,+1)$\\
$m_7   $&  $(-1,-1)$ & $(-1,-1)$ &$(-1,-1)$\\
\hline\noalign{\smallskip}
\end{tabular}
\label{tab4}
\end{table}

\begin{table}[!htp]
\caption{
Total counts of different pairs $(A,B)$ belonging to the set of triples.
}
\centering
\begin{tabular}{@{\extracolsep{.2cm} } cccc}
\noalign{\medskip}
\hline\hline\noalign{\smallskip}
$(A,B)$&$n_1(A_1,B_1)$&$n_2(A_2,B_2)$&$n_3(A_3,B_3)$\\
\hline\noalign{\smallskip}
$(+1,+1)$&$m_0+m_2$ & $m_0+m_1$ & $m_0+m_4$ \\
$(+1,-1)$&$m_1+m_3$ & $m_2+m_3$ & $m_2+m_6$ \\
$(-1,+1)$&$m_4+m_6$ & $m_4+m_5$ & $m_1+m_5$ \\
$(-1,-1)$&$m_5+m_7$ & $m_6+m_7$ & $m_3+m_7$ \\
\hline\noalign{\smallskip}
\end{tabular}
\label{tab5}
\end{table}

Next, count the number of times a pair $(\pm1,\pm1)$ occurs in the data sets ${\cal D}_{\Cac}$, ${\cal D}_{\Cad}$,
and ${\cal D}_{\Cbc}$,
and denote the twelve numbers thus obtained by $N_1(+1,+1),N_1(-1,+1),\ldots, N_3(+1,-1),N_3(-1,-1)$.
These numbers completely determine the values of the correlations
$C_{\Cac}$, $C_{\Cad}$, and $C_{\Cbc}$,
e.g., $C_{\Cac}=(N_1(+1,+1)-N_1(+1,-1)-N_1(-1,+1)+N_1(-1,-1))/N$.
The same twelve numbers serve as input to the linear optimization problem to be formulated next.

Denoting the number of pairs $(A,B)$ in data set ${\cal D}_k$ that do
not belong to the set of triples by $u_k(A,B)\ge0$, we must have
\begin{eqnarray}
N_k(A,B)=n_k(A,B)+u_k(A,B)
\;,\;
N=\sum_{x,y=\pm1} N_k(x,y)
\;,\;
U=\sum_{x,y=\pm1} u_k(x,y)
\;,
\label{QUA0}
\end{eqnarray}
for $k=1,2,3$ and all pairs $(A,B)=(\pm1,\pm1)$.
Note that $U$ cannot depend on $k$ because the number of pairs
which do not belong to the set of triples must be the same for all three data sets.

The final step is then to minimize the number of pairs which do not belong
to the set of triples, by solving the linear minimization problem
\begin{equation}
  \min\left(U=N-\sum_{i=0}^{7} m_i\right)
\end{equation}
in 20 unknowns (the eight $m_i$'s and twelve $u_k(A,B)$'s),
subject to 21 inequality constraints ($m_i,u_k(\pm1,\pm1),U\ge0$ for all $i,k$) and 14
equality constraints (see Eq.~(\ref{QUA0})).

At first sight, finding the value of $\TRIP$ seems to require ${\cal O}(N!^2)$ operations.
Fortunately, the problem of determining the fraction of triples $\TRIP$ can
be cast into an integer linear programming problem which is readily solved
by considering the associated linear programming problem with real-valued
unknowns. In practice, we solve the latter by standard optimization
techniques~\cite{PRES03}.
The computer program to calculate the maximum fraction of triples
has been implemented in Mathematica\textsuperscript{\textregistered}.
For all cases that have been explored, the solution of
the linear programming problem takes integer values only. Then the
solution of the linear programming problem is also the solution of the integer
programming problem.

It is instructive to perform numerical experiments with synthetic data.
The results of such experiments using $N=1000 000$ pairs per data set ${\cal D}_s$ can be summarized as follows:
\begin{enumerate}
\item
If all $A$'s and $B$'s take independent random values $\pm1$,
the program returns $C_{\Cac}=-0.001118$, $C_{\Cad}= -0.001954$,
$C_{\Cbc}=-0.000088$, and
$\widetilde\TRIP\approx0.99932$.
This, perhaps counter intuitive observation that in this particular case,
it is possible to reshuffle almost all data to form triples may be understood as follows.
If $N$ is very large, it is (in the case at hand) to be expected that
$N_1(+1,+1)\approx N_1(+1,-1)\approx\ldots\approx N_3(+1,-1)
\approx N_3(-1,-1)\approx {N/4}$.
It then follows that $\TRIP=1-\epsilon$ where,
in the numerical experiment, $\epsilon=0.00068$ gives an indication of
the statistical fluctuations.
\item
If the pairs $(A_{\Cac,i},B_{\Cac,i})$,
$(A_{\Cad,j},B_{\Cad,j})$, and
$(A_{\Cbc,k},B_{\Cbc,k})$
are generated randomly with frequencies
$(1-c_{\Cac}A_{\Cac,i}B_{\Cac,i})/4$,
$(1-c_{\Cad}A_{\Cad,j}B_{\Cad,j})/4$, and
$(1-c_{\Cbc}A_{\Cbc,j}B_{\Cbc,j})/4$, respectively,
the simulation can mimic the case
of the correlation of two spin-1/2 objects in the singlet state.
Choosing $c_{\Cac}=-c_{\Cad}=1/\sqrt{2}$ and $c_{\Cbc}=0$, quantum theory yields
$|C_{\Cac}- C_{\Cad}|=\sqrt{2}\approx1.41$ and $C_{\Cbc}=0$~\cite{RAED23}.
Generating independent pairs, the program returns
$|C_{\Cac}-C_{\Cad}|=1.41384$,
$C_{\Cbc}=0.000432$ and $3-2\TRIP-C_{\Cbc}=1.41384$, 
demonstrating that the value of the quantum-theoretical upper bound~\cite{CIRE80} of $\sqrt{2}$
is reflected in the maximum fraction of triples that
one can find by reshuffling the data.
In this case the data suggest that inequality Eq.~(\ref{DISD9}) is saturated.
\end{enumerate}

Note that the numerical values given in 1--2 are subject to minute changes stemming
from the use of the pseudo-random number generator provided by Mathematica\textsuperscript{\textregistered}.

\section{Basic inequalities}\label{BASIC}

For any pair of real numbers $u$ and $v$, the triangle inequality
$|u+v|\le |u|+|v|$
and the identity
$(u \pm v)^2+(1-u^2)(1-v^2)=(1\pm uv)^2$
hold.
If the symbols $x$ and $y$ represent real numbers in the range $[-1,1]$,
the second term on the left hand side of the latter identity with $u$ and $v$ replaced
by $x$ and $y$, respectively, is nonnegative.
This implies that
$(x \pm y)^2\le (1\pm xy)^2$ or, equivalently, $|x \pm y|\le 1\pm xy$.
If the symbol $z$ also represents a real number in the range $[-1,1]$,
using the triangle inequality it immediately follows that
\begin{eqnarray}
|xz \pm yz| &=&|z||x\pm y|\le1\pm xy
\;.
\label{BASIC3}
\end{eqnarray}
The variables appearing in Eq.~(\ref{BASIC3}) form the triple $(x,y,z)$,
a structure which is essential to prove Eq.~(\ref{BASIC3}).

It is straighforward to show (see Appendix N of Ref.~\cite{RAED23}) that for any three real numbers $a$, $b$, and $c$,
\begin{eqnarray}
|a\pm b|\le 1 \pm c \iff |a\pm c|\le 1 \pm b \iff |b\pm c|\le 1 \pm a
\;,
\label{TRIPLE}
\end{eqnarray}
equivalences that are useful to generate ``new'' inequalities.

\fi

\section{Trivariates of two-valued variables}\label{FINE}

Without loss of generality, any real-valued, normalized function $f(x_1,x_2,x_3)$ of the two-valued variables
$x_1=\pm1$, $x_2=\pm1$, and $x_3=\pm1$ can be written as
\begin{eqnarray}
f(x_1,x_2,x_3)&=&\frac{1 + K_{1}\,x_1 + K_{2}\,x_2 + K_{3}\,x_3+ K_{12}\,x_1x_2
+ K_{13}\,x_1x_3 + K_{23}\,x_2x_3 + K_{123}\,x_1x_2x_3 }{8}
\;.
\label{LGI0}
\end{eqnarray}
From Eq.~(\ref{LGI0}) it follows that
\begin{subequations}
\label{LGI1}
\begin{eqnarray}
1&=&\sum_{x_1=\pm1}\sum_{x_2=\pm1}\sum_{x_3=\pm1} f(x_1,x_2,x_3)
\;,
\label{LGI1a}
\\
K_{i}&=&\sum_{x_1=\pm1}\sum_{x_2=\pm1}\sum_{x_3=\pm1} x_i f(x_1,x_2,x_3)\;,\; i\in\{1,2,3\}
\;,
\label{LGI1b}
\\
K_{ij}&=&\sum_{x_1=\pm1}\sum_{x_2=\pm1}\sum_{x_3=\pm1} x_ix_j f(x_1,x_2,x_3)
\;,\; (i,j)\in\{(1,2),(1,3),(2,3)\}
\;,
\label{LGI1c}
\\
K_{123}&=&\sum_{x_1=\pm1}\sum_{x_2=\pm1}\sum_{x_3=\pm1} x_1x_2x_3 f(x_1,x_2,x_3)
\;,
\label{LGI1d}
\end{eqnarray}
\end{subequations}
where the $K$'s are the moments of $f(x_1,x_2,x_3)$ and Eq.~(\ref{LGI1a})
is a restatement of the normalization of $f(x_1,x_2,x_3)$.

If $f(x_1,x_2,x_3)$ is going to be used as a model for empirical frequencies,
it must satisfy $0\le f(x_1,x_2,x_3) \le 1$ and $\sum_{x_1,x_2,x_3=\pm1} f(x_1,x_2,x_3)=1$.
From $0\le f(x_1,x_2,x_3)\le 1$, it follows immediately that all the $K$'s in
Eq.~(\ref{LGI1}) are smaller than one in absolute value.
Furthermore the marginals
$f_3(x_1,x_2)=\sum_{x_3=\pm1}f(x_1,x_2,x_3)$,
$f_2(x_1,x_3)=\sum_{x_2=\pm1}f(x_1,x_2,x_3)$, and
$f_1(x_2,x_3)=\sum_{x_1=\pm1}f(x_1,x_2,x_3)$
are real-valued, normalized and nonnegative bivariates
that is $0\le f_3(x_1,x_2) \le 1$, $\sum_{x_1,x_2=\pm1} f_3(x_1,x_2)=1$, etc.
From the nonnegativity of these marginals, it follows that
$|K_i\pm K_j|\le 1 \pm K_{ij}$ for $(i,j)\in\{(1,2),(1,3),(2,3)\}$~\cite{RAED23}.
Other inequalities involving moments follow by making linear combinations of the inequalities $f(x_1,x_2,x_3)\ge0$
for different values of $(x_1,x_2,x_3)$.
For instance, from
$4[f(+1,+1,+1) + f(-1,-1,-1)] = 1+K_{12}+K_{13}+K_{23} \ge 0$
and
$4[f(-1,+1,+1) + f(+1,-1,-1)] = 1-K_{12}-K_{13}+K_{23} \ge 0$
it follows that $|K_{12}+K_{13}|\le 1 + K_{23}$, one instance the Boole-Bell inequality.
Recall that the latter implies that the inequalities
$\vert K_{12}\pm K_{23}\vert \le 1 \pm K_{13}$ and
$\vert K_{13}\pm K_{23}\vert \le 1 \pm K_{12}$ are also satisfied
\REP{(see Eq.~(\ref{TRIPLE}))}{see Eq.~(N.6) in Ref.~\cite{RAED23}}.

Summarizing: if the data can be modeled by
a nonnegative, normalized trivariate $f(x_1,x_2,x_3)$, all inequalities
\begin{subequations}
\label{THREE3}
\begin{align}
|K_{1}|&\le1\;,\; |K_{2}|\le1 \;,\;|K_{3}|\le1\;,
|K_{12}|\le1\;,|K_{13}|\le1\;,\;|K_{23}|\le1
\;,
\label{THREE3a}
\\
\vert K_{1}\pm K_{2}\vert &\le 1 \pm K_{12}\;,\;
\vert K_{1}\pm K_{3}\vert \le 1 \pm K_{13}\;,\;
\vert K_{2}\pm K_{3}\vert \le 1 \pm K_{23}\;,\;
\label{THREE3b}
\\
\vert K_{12}\pm K_{13}\vert &\le 1 \pm K_{23}\;,\;
\label{THREE3c}
\end{align}
\end{subequations}
which include the Boole-Bell inequalities are satisfied.
Thus, the inequalities Eq.~(\ref{THREE3}) do not only derive from Bell's model Eq.~(\ref{IN0}) but
also from a much more general model defined by the trivariate Eq.~(\ref{LGI0}).

A remarkable fact, first shown by Fine~\cite{FINE82a,FINE82b} by a different approach than the one taken here,
is that if all inequalities Eq.~(\ref{THREE3})
are satisfied, it is possible to construct a real-valued,
normalized trivariate $0\le f(x_1,x_2,x_3) \le 1$
of the two-valued variables $x_1=\pm1$, $x_2=\pm1$, and $x_3=\pm1$,
yielding all the moments that appear in Eq.~(\ref{THREE3}).
The text that follows replaces the corresponding part and theorem in Ref.~\cite{RAED23}, which are not correct.

First note that if Eqs.~(\ref{THREE3a}) and~(\ref{THREE3b}) hold,
the existence of the three normalized bivariates $0\le f_3(x_1,x_2) \le 1$,
$0\le f_2(x_1,x_3) \le 1$, and $0\le f_1(x_2,x_3) \le 1$ is guaranteed~\cite{RAED23}.
Indeed, for instance, if
$|K_{1}|\le1$, $|K_{2}|\le1$, $|K_{12}|\le1$, and $\vert K_{1}\pm K_{2}\vert \le 1 \pm K_{12}$
it follows immediately that
$0\le f_3(x_1,x_2)=(1 + K_{1}\,x_1 + K_{2}\,x_2 + K_{12}\,x_1x_2)/4\le1$
is the desired normalized bivariate.
Therefore, what remains to be proven is the existence of a real-valued trivariate $g(x_1,x_2,x_3)$ that
(i) takes values in the interval $[0,1]$ and (ii)
yields the three named bivariates with their respective moments $K_1,\ldots,K_{23}$ that appear
in Eqs.~(\ref{THREE3a}) and~(\ref{THREE3b}) as marginals.

Second, note that without loss of generality, any real-valued trivariate $g(x_1,x_2,x_3)$ of the two-valued variables
$x_1=\pm1$, $x_2=\pm1$, and $x_3=\pm1$ can be written as

\begin{eqnarray}
g(x_1,x_2,x_3)&=&\frac{K'_0 + K'_{1}\,x_1 + K'_{2}\,x_2 + K'_{3}\,x_3+ K'_{12}\,x_1x_2
+ K'_{13}\,x_1x_3 + K'_{23}\,x_2x_3 + K'_{123}\,x_1x_2x_3 }{8}
\;.
\label{g0}
\end{eqnarray}
Imposing requirement (ii) immediately yields $K'_0=1$,
$K'_{1}=K_{1}$,
$K'_{2}=K_{2}$,
$K'_{3}=K_{3}$,
$K'_{12}=K_{12}$,
$K'_{13}=K_{13}$, and
$K'_{23}=K_{23}$,
leaving only $K'_{123}$ as unknown.

Third, the requirement that $0\le g(x_1,x_2,x_3)$ is used to
derive conditions for the unknown $K'_{123}$ in terms of all the moments that appear in Eq.~(\ref{THREE3}).
This can be accomplished as follows.
The eight inequalities $g(x_1,x_2,x_3)\ge0$ are rewritten as bounds on $K'_{123}$.
For instance
$g(+1,+1,+1)\ge0 \iff -1 - K_{1} - K_{2} - K_{3}-  K_{12}- K_{13} - K_{23}\le K'_{123}$
and $g(-1,-1,-1)\ge0 \iff K'_{123}\le 1 - K_{1} - K_{2} - K_{3} + K_{12}+ K_{13} + K_{23}$,
and so on.
The resulting eight inequalities can be summarized as
\begin{align}
\max\big[&-1+
K_1-K_2-K_3-K_{12}-K_{13}+K_{23},
-1-K_1+K_2-K_3-K_{12}+K_{13}-K_{23},\nonumber \\
&-1-K_1-K_2+K_3+K_{12}-K_{13}-K_{23},
-1+K_1+K_2+K_3+K_{12}+K_{13}+K_{23}
\big]\nonumber \\
&
\hbox to -.5cm{}\le K_{123}\le
\min\big[
1-K_1-K_2-K_3+K_{12}+K_{13}+K_{23},
1+K_1+K_2-K_3+K_{12}-K_{13}-K_{23},\nonumber \\
&\hbox to 1.72cm{}
1+K_1-K_2+K_3-K_{12}+K_{13}-K_{23},
1-K_1+K_2+K_3-K_{12}-K_{13}+K_{23}
\big]
\;.
\label{K123}
\end{align}
Using the inequalities $\max(a,b,c,d)\ge (a+b+c+d)/4$ and $\min(a,b,c,d)\le (a+b+c+d)/4$
it immediately follows from Eq.~(\ref{K123}) that $-1\le K'_{123}\le1$,
which together with Eq.~(\ref{THREE3a}), guarantees that $g(x_1,x_2,x_3)\le1$.

To prove that there exists at least one value of $K'_{123}$ satisfying Eq.~(\ref{K123})
if all inequalities in Eq.~(\ref{THREE3}) are satisfied, it is sufficient to show that the difference
$\min(\ldots) - \max(\ldots)$ cannot be negative.
One straighforward way to do this is to
use the sixteen inequalities Eqs.~(\ref{THREE3b}) and Eq.~(\ref{THREE3c}) to show
that all sixteen possible differences deriving from Eq.~(\ref{K123}) are nonnegative.
Instead, it is more elegant to rewrite the arguments of $\max(\ldots)$ and $\min(\ldots)$ in Eq.~(\ref{K123}) as
\begin{subequations}
\label{g2}
\begin{align}
\label{g2a}
  K'_{123} &\ge \max( -1-K_3-K_{12}+| K_1+K_2+K_{13}+K_{23} |, -1+K_3+K_{12}+| K_1-K_2-K_{13}+K_{23} | ) & =:&\ \LHS\;,\\
\label{g2b}
  K'_{123} &\le \min( 1-K_3+K_{12}-| K_1+K_2-K_{13}-K_{23} |,  1+K_3-K_{12}-| K_1-K_2+K_{13}-K_{23} | ) &  =:&\ \RHS\;.
\end{align}
\end{subequations}
The final step is then to prove that $\RHS-\LHS\ge0$.
Combining Eqs.~(\ref{g2a}) and~(\ref{g2b}) and using $\min(a,b)+\min(c,d)=\min(a+c,a+d,b+c,b+d)$ yields
\begin{align}
  \RHS - \LHS =
\min (&2+2K_{12}-|K_1+K_2-K_{13}-K_{23}|-|K_1+K_2+K_{13}+K_{23}|\;,\nonumber\\
 &2-2K_{12}-|K_1-K_2+K_{13}-K_{23}|-|K_1-K_2-K_{13}+K_{23}|\;, \nonumber\\
 &2-2K_3-|K_1+K_2-K_{13}-K_{23}|-|K_1-K_2-K_{13}+K_{23}|\;, \nonumber\\
 &2+2K_3-|K_1-K_2+K_{13}-K_{23}|-|K_1+K_2+K_{13}+K_{23}|)
 \;.
 \label{g4}
\end{align}
Using the inequalities Eq.~(\ref{THREE3}) and the identity  $-|a-b|-|a+b| = \min(-a+b,-b+a)+\min(-a-b,a+b)= 2\min(-|a|,-|b|)$,
it can be shown that each of the four arguments of $\min(.)$ in Eq.~(\ref{g4}) is nonnegative.
In detail
\begin{subequations}
\begin{align}
 2+2K_{12}-|K_1+K_2-(K_{13}+K_{23})|-|K_1+K_2+K_{13}+K_{23}| &= 2+2K_{12}+2\min(-|K_1+K_2|,-|K_{13}+K_{23}|)\nonumber\\
 &\ge 2+2K_{12} +2(-1-K_{12}) = 0\; ,\\
 2-2K_{12}-|K_1-K_2+K_{13}-K_{23}|-|K_1-K_2-(K_{13}-K_{23})| &= 2-2K_{12}+2\min(-|K_1-K_2|, -|K_{13}-K_{23}|)\nonumber\\
 &\ge 2-K_{12}+2(-1+K_{12}) = 0\; ,\\
 2-2K_3-|K_1-K_{13}+K_2-K_{23}|-|K_1-K_{13}-(K_2-K_{23})| &= 2-2K_3+2\min(-|K_1-K_{13}|,-|K_2-K_{23}|) \nonumber\\
 &\ge 2-2K_3+2(-1+K_3) = 0\; ,\\
 2+2K_3-|K_1+K_{13}-(K_2+K_{23})|-|K_1+K_{13}+K_2+K_{23}| &= 2+2K_3+2\min(-|K_1+K_{13}|,-|K_2+K_{23}|)\nonumber\\
 &\ge 2+2K_3+2(-1-K_3) = 0\; .
\end{align}
\end{subequations}
This completes the proof that if all inequalities Eq.~(\ref{THREE3}) are satisfied,
there exists a normalized, nonnegative trivariate $0\le g(x_1,x_2,x_3)\le1$ given by
Eq.~(\ref{g0}) with $K'_{123}\in[\LHS,\RHS]$.
Summarizing we have proven

\newsavebox\THEOREM
\savebox\THEOREM{{\bf Theorem:\ }} 
\begin{center}
\framebox{
\parbox[t]{0.85\hsize}{%
\hangindent=\wd\THEOREM
\hangafter=1
\usebox\THEOREM
Given a real-valued, normalized function
$0\le f(x_1,x_2,x_3)\le 1$ of two-valued variables, its moments Eq.~(\ref{LGI1a})--(\ref{LGI1c})
satisfy all the inequalities Eq.~(\ref{THREE3}).
Conversely,
given the values of the moments Eq.~(\ref{LGI1a})--(\ref{LGI1c})
satisfying all the inequalities Eq.~(\ref{THREE3}), it is always possible to choose $K_{123}$ in the range $[\LHS,\RHS]$
(defined by the right hand sides of Eqs.~(\ref{g2a}) and~(\ref{g2b}), respectively),
and construct a real-valued, normalized function $0\le f(x_1,x_2,x_3)\le 1$ of two-valued variables
which yields the specified values of the moments Eq.~(\ref{LGI1a})--(\ref{LGI1c}).
}}
\end{center}
A different strategy was implemented in Mathematica\textsuperscript{\textregistered},
providing an independent proof of the theorem.
The explicit form of $f(x_1,x_2,x_3)$ in terms of its moments is given by Eq.~(\ref{LGI0}).


\bibliographystyle{elsarticle-num}
\bibliography{/D/papers/all24}

\end{document}